\documentclass[aps,floats,amssymb,amsmath,prb,twocolumn,superscriptaddress,longbibliography]{revtex4-1}
\usepackage{graphicx}
\usepackage[english]{babel}
\usepackage{times}
\usepackage{color}
\usepackage{bm}
\usepackage{capt-of}
\usepackage{braket}
\usepackage{float}
\usepackage{epstopdf}
\usepackage{hyperref}
\usepackage{xcolor}
\hypersetup{
    colorlinks,
    linkcolor={red!50!black},
    citecolor={blue!50!black},
    urlcolor={blue!80!black}
}

\DeclareMathAlphabet\mathbfcal{OMS}{cmsy}{b}{n}

\newcommand{\veck}{\mathbf{k}}

\newcommand{\Dbcs}{\Delta_{\mathrm{BCS}}}

\newcommand{\PSI}{\mathbf{\Psi}}
\newcommand{\wn}{\omega_n}

\begin{document}
\title{Reentrant $s$-wave superconductivity in the periodic Anderson model with attractive conduction band Hubbard interaction}

\author{W.-V. van Gerven Oei}
\affiliation{Institute of Physics Belgrade, University of Belgrade, Pregrevica 118, 11080 Belgrade, Serbia}

\author{D. Tanaskovi\'c}
\affiliation{Institute of Physics Belgrade, University of Belgrade, Pregrevica 118, 11080 Belgrade, Serbia}

\begin{abstract}

Spin-flip scattering from magnetic impurities has a strong pair-breaking effect in $s$-wave superconductors where increasing the concentration of impurities rapidly destroys superconductivity. 
For small Kondo temperature $T_K$ the destruction of superconductivity is preceded by the reentrant superconductivity at finite temperature range $T_{c2} < T < T_{c1}$, while the normal phase reappears at $T<T_{c2} \sim T_K$. Here we explore the superconducting phase in a periodic system modeled as the Anderson lattice with additional attractive on-site (Hubbard) interaction $g$ acting on the conduction band electrons. We solve the equations using dynamical mean field theory which incorporates Kondo physics, while the pairing interaction is treated on the static mean-field level. For large coupling $g$ we find reentrant superconductivity which resembles the case with diluted impurities. However, we find evidence that reentrant superconductivity is here not a consequence of many-body correlations leading to the Kondo effect,
but it rather stems from a competition between the single-particle hybridization and superconducting pairing. An insight into the spectral functions with in-gap structures is obtained from an approximate noninteracting dual model whose solution interpolates between several exact limits.

\end{abstract}


\maketitle

\section{Introduction}

The influence of magnetic impurities on conventional $s$-wave superconductors has been intensively explored since early sixties\cite{AG_JETP1961,BalatskyRMP2006,Wolowiec_review2015,Bauer_2007}.
In a seminal work Abrikosov and Gor'kov (AG) have shown\cite{AG_JETP1961}, within the second-order Born approximation, that the scattering from impurity spins breaks the Cooper pairs and suppresses superconductivity. AG theory predicts the decrease in transition temperature $T_c$ determined by a universal function of the pair-breaking parameter which is proportional to the impurity concentration, spin magnitude, and the exchange interaction. There is excellent quantitative agreement between the AG theory and numerous experiments on conventional superconductors with rare earth impurity ions\cite{Wolowiec_review2015,Maple_1968,Fisk_1969}. 

A notable deviation from the AG theory is observed in some alloys, like La$_{1-x}$Ce$_x$Al$_2$\cite{Maple_1972,Riblet_1972}. In these systems $T_c$ initially decreases with increasing the concentration $x$ similar as in AG theory, but near the critical concentration there is a regime where the system is superconducting below an ``upper'' critical temperature $T_{c1}$, but it ``reenters'' the normal phase at nonzero $T_{c2}$. This reentrant superconductivity is explained as a consequence of Kondo physics\cite{MHZ1970,Schlottmann_1975}, and it appears when the characteristic Kondo temperature $T_K$ is much smaller than the critical temperature of the clean system $T_{c0}$. 
The impurity scattering and the pair-breaking parameter acquire strong temperature dependence. At temperatures $T \gg T_K$ the impurity spins are weakly coupled to the conduction electrons and the superconducting (SC) phase persists, while the normal phase reappears at temperatures $T\sim T_K$ when the scattering becomes stronger. 
More recently, reentrant superconductivity is obtained from a solution of the Eliashberg equations supplemented by the quantum Monte Carlo solution of the Anderson impurity problem\cite{Jarrell90b}. The critical concentration for the full suppression of the superconductivity grows with increasing the electron-phonon coupling, but it typically remains of the order of 1\%.

Can the superconductivity and the reentrant behavior persist in the case of periodic impurities, i.e.~in the presence of a second band of interacting dispersionless electrons hybridized with the conduction band? 
The reentrant superconductivity is indeed observed in several ternary\cite{Ishikawa_1977,Moncton_77,Fertig_1977,Remeika_1980} and quaternary conventional superconductors\cite{Eisaki_1994} with periodic weakly hybridized rare-earth magnetic ions. The reentrant behavior is here attributed to the magnetic ordering for $T<T_{c2}$ and not the Kondo physics. More recently, the reentrant superconductivity is observed also in some iron based superconductors like EuFe$_2$As$_2$\cite{Miclea_PRB2009,Paramanik_2013}.

In heavy fermion compounds the superconductivity is mediated by spin fluctuations instead of the electron-phonon coupling\cite{Steglich_2016}. This generically leads to $d$-wave pairing, though recent studies on $\text{CeCu}_2\text{Si}_2$ surprisingly indicated a fully gapped $s$-wave state\cite{Takenaka_PRL2017}. 
Hence, theoretical work on the systems with periodic magnetic moments was mostly focused on unconventional $d$-wave pairing near the antiferromagnetic quantum critical point in the Kondo/Anderson lattice model\cite{Wu_PRX2014,Asadzadeh_PRB2014,Otsuki_PRL2015}, with few exceptions that treated attractive on-site pairing interaction in the Kondo lattice model \cite{Lechtenberg_PRB2018,Costa2018,Bodensiek_2010,Bertussi_PRB2009}.

In this work we explore the effect of periodic magnetic impurities on conventional superconductivity. Our starting point is the Anderson lattice model with the addition of an on-site pairing (attractive Hubbard) interaction acting on the conduction $c$-electrons. The repulsive interaction $U$ on $f$-orbitals is treated within dynamical mean field theory\cite{Georges96} (DMFT) using continuous time hybridization expansion quantum Monte Carlo (Cthyb QMC) impurity solver\cite{Haule_PRB2007}, while the on-site pairing $g$ is treated on the static mean-field level. This model is closely related to the Kondo lattice model that has been very recently studied\cite{Lechtenberg_PRB2018,Costa2018}, but here we focus on finite temperatures and away from half-filling where magnetic and charge density wave instabilities are expected to be weaker. We study the superconductivity phase diagram for different pairing couplings $g$ and hybridization $V$. For strong coupling $g$ we find reentrant superconductivity which resembles the one seen in the diluted impurities case. In the weak coupling case we could not identify if the reentrance persists due to the very small relevant energy scales that cannot be accessed by the QMC solver.
We have also solved the model for parameters away from the Kondo limit and found that the reentrant superconductivity may appear in some cases due to band structure physics, i.e.~due to the competition between single-particle hybridization and superconducting pairing. In order to better understand the electronic spectrum of the model Hamiltonian, we have also introduced and solved an approximate noninteracting dual model\cite{Sakai_PRB2016,Sakai_PRL2016}.

The paper is organized as follows. Section II contains the definition of the model and describes the methods of its solution. Numerical results in several parameter regimes are shown in Section III and our conclusions are in Section IV. Some derivations are presented in the Appendix.

\section{Model and methods}

We solve the periodic Anderson model with an additional attractive Hubbard interaction in the conduction band. The Hamiltonian is given by
\begin{align}\label{eq:hamiltonian}
H = -&t\sum_{\braket{ij}\sigma}(c^\dagger_{i\sigma}c_{j\sigma} + \mathrm{H. c.}) - \mu  \sum_{i\sigma}c^\dagger_{i\sigma}c_{i\sigma}  \nonumber\\
& -V\sum_{i\sigma}(c^\dagger_{i\sigma}f_{i\sigma} + \mathrm{H. c.}) + (\epsilon_f - \mu) \sum_{i\sigma}f^\dagger_{i\sigma}f_{i\sigma}\nonumber \\
& - g\sum_i c^\dagger_{i\uparrow}c^\dagger_{i\downarrow}c_{i\downarrow}c_{i\uparrow} + U\sum_i f^\dagger_{i\uparrow}f^\dagger_{i\downarrow}f_{i\downarrow}f_{i\uparrow},
\end{align}
where $t$ is the hopping parameter, $V$ the hybridization strength, $g$ is the attractive coupling for the conduction band $c$-electrons, and $U$ is the repulsive coupling constant of the $f$-electrons. $\epsilon_f$ sets the energy level of the $f$-electrons, and $\mu$ is the chemical potential. $c_{i\sigma}^\dagger$ and $f_{i\sigma}^\dagger$ create a  $c$-electron and $f$-electron at site $i$ with spin $\sigma=\uparrow,\downarrow$.
This model reduces to the attractive Hubbard model with decoupled impurities in the limit $V\rightarrow0$, whereas in the limit $g\rightarrow0$ we recover the standard Anderson lattice model. 
We take as a unit of energy the half-bandwidth $D$ corresponding to the noninteracting $c$ electrons. We will restrict to the paramagnetic solution allowing for $s$-wave superconductivity. 

We start with a static mean-field decoupling of the $c$-electron attractive interaction in the Cooper channel, viz.
\begin{equation}\label{eq:Cooper_channel}
g\sum_i c^\dagger_{i\uparrow}c^\dagger_{i\downarrow}c_{i\downarrow}c_{i\uparrow} \rightarrow \Dbcs\sum_i(c^\dagger_{i\uparrow}c^\dagger_{i\downarrow} + \mathrm{H.c.}) ,
\end{equation}
where $\Dbcs = g\braket{c^\dagger_{i\uparrow}c^\dagger_{i\downarrow}} = g\braket{c_{i\downarrow}c_{i\uparrow}} = g \Phi_c$ is the superconducting order parameter. 
This recasts the problem in the form of a self-consistently determined Hamiltonian $H[\Delta_{\mathrm{BCS}}]$, featuring both the pairing terms and the repulsive Hubbard interaction. Without the repulsive Hubbard interaction, this reduces to a Bardeen-Cooper-Schrieffer (BCS) mean-field theory, hence the index in the pairing amplitude.

We now introduce momentum-dependent fermionic Grassmann fields in orbital-Nambu space:
\begin{align}\label{eq:tau_fields}
&\PSI_\veck(\tau) = \begin{bmatrix}
	\mathbf{c}_\veck(\tau) \\
	\mathbf{f}_\veck(\tau)
\end{bmatrix},\nonumber\\
&\mathbf{c}_\veck(\tau) = \begin{pmatrix}
	c_{\veck\uparrow}(\tau) \\
	\bar{c}_{-\veck\downarrow}(\tau)
\end{pmatrix} ,\;\;\;\mathbf{f}_\veck(\tau) = \begin{pmatrix}
	f_{\veck\uparrow}(\tau) \\
	\bar{f}_{-\veck\downarrow}(\tau)
\end{pmatrix}.
\end{align}
Here $\tau$ is the imaginary time variable and overbar indicates the conjugate field. In Grassmann field formalism, the action for the self-consistent Hamiltonian reads
\begin{align}\label{eq:action}
S =-&\int_0^\beta d\tau\int_0^\beta d\tau'\sum_{\veck}\bar{\PSI}_\veck(\tau)\mathbf{G}_{0,\veck}^{-1}(\tau-\tau')\PSI_\veck(\tau') \nonumber\\
& + U\int_0^\beta d\tau\sum_{\veck,\mathbf{k}',\mathbf{q}} \bar{f}_{\veck+\mathbf{q}\uparrow}(\tau)\bar{f}_{\veck'-\mathbf{q}\downarrow}(\tau)f_{\veck'\downarrow}(\tau)f_{\veck\uparrow}(\tau).
\end{align}
$\beta$ is the inverse temperature and $\mathbf{G}_{0,\veck}$ is the bare propagator, implicitly dependent on $\Dbcs$. 
In Matsubara frequency domain, the bare propagator reads
\begin{equation}\label{eq:G0}
\mathbf{G}_{0,\veck}(i\wn) = \left[i\wn\mathbf{I} - \mathbf{H}_{0,\veck}\right]^{-1}.
\end{equation}
Here $\wn$ are fermionic Matsubara frequencies $\wn=(2n+1)\pi/\beta$, $\mathbf{I}$ is the 4-dimensional identity matrix, and $\mathbf{H}_{0,\veck}$ is the non-interacting Hamiltonian matrix in the orbital-Nambu basis, i.e.
\begin{equation}\label{eq:H0}
\mathbf{H}_{0,\veck} = \begin{pmatrix}
	\xi_\veck & -\Dbcs & -V & 0 \\
	-\Dbcs & -\xi_\veck & 0 & V \\
	-V & 0 & \epsilon_f-\mu & 0 \\
	0 & V & 0 & -\epsilon_f+\mu
\end{pmatrix},
\end{equation}
where $\xi_\veck \equiv \varepsilon_\veck - \mu$.

The full (interacting) Green's function in the Matsubara domain is defined component-wise as
\begin{align}\label{eq:def_Gk}
\mathbf{G}_\veck = -\braket{\PSI_\veck\otimes\PSI^\dagger_\veck} &\equiv \begin{pmatrix}
	G_{c,\veck} & \mathcal{F}_{c,\veck} & G_{cf,\veck} & \mathcal{F}_{cf,\veck} \\
	\mathcal{F}_{c,\veck} & -G^*_{c,\veck} & \mathcal{F}_{cf,\veck} & G_{cf,\veck} \\
	-G^*_{cf,\veck} & \mathcal{F}_{cf,\veck} & G_{f,\veck} & \mathcal{F}_{f,\veck} \\
	\mathcal{F}_{cf,\veck} & -G^*_{cf,\veck} & \mathcal{F}_{f,\veck} & -G^*_{f,\veck}
\end{pmatrix} \nonumber \\
&\equiv \begin{bmatrix}
	\mathbf{G}_{c,\veck} & \mathbf{G}_{cf,\veck} \\
	\mathbf{G}_{fc,\veck} & \mathbf{G}_{f,\veck}
\end{bmatrix}.
\end{align}
where we have used $G_{c/f,\veck}(-i\omega_n)=G_{c/f,\veck}^*(i\omega_n)$, and the lattice inversion symmetry $\veck\rightarrow -\veck$.
The second equivalence states the definitions of the $c$ and $f$ Nambu (two-dimensional) Green's functions in their respective orbital subsectors, and the $\wn$-dependence is implicit.

The full $\mathbf{G}_\veck$ is to be determined through the Dyson equation
\begin{equation}\label{eq:Gk}
\mathbf{G}_\veck^{-1}(i\wn) = \mathbf{G}_{0,\veck}^{-1}(i\wn) - \mathbf{\Sigma}_\veck(i\wn)
\end{equation}
where $\mathbf{\Sigma}$ is the matrix self-energy capturing the on-site correlation effects, viz.
\begin{equation}\label{eq:Sigma}
\mathbf{\Sigma}_\veck = \begin{pmatrix}
	0 & 0 & 0 & 0 \\
	0 & 0 & 0 & 0 \\
	0 & 0 & \Sigma_{\veck} & \mathcal{S}_\veck \\
	0 & 0 & \mathcal{S}_\veck & -\Sigma^*_{\veck}
\end{pmatrix}
.
\end{equation}
$\mathcal{S}_\veck$ is the self-energy's anomalous component, and satisfies $\mathcal{S}_\veck(i\wn\rightarrow\infty)=U\mathcal{F}_{f,\veck}(\tau=0)$. The superconducting order parameter is determined from the scalar $c$-electrons' Green's function as 
\begin{align}\label{eq:order_param}
\Dbcs &= \frac{g}{N_k}\sum_\veck\mathcal{F}_{c,\veck}(\tau=0).
\end{align}
Here $N_k$ is the total number of momenta in the discretized first Brillouin zone. 
Henceforth, the local quantities will be indicated by omitting the $\veck$ index, while the normalization constant $N^{-1}_k$ will be absorbed into the sum---e.g. Eq. \eqref{eq:order_param} then reads $\Dbcs = g\mathcal{F}_c(\tau=0)$.

\subsection{DMFT}

We solve the self-consistent problem Eq.(\ref{eq:action}) using the dynamical mean field theory.
DMFT assumes the self-energy to be entirely local, i.e. $\Sigma_\veck\rightarrow\Sigma$. The local self energy is computed from an effective single-impurity problem
\begin{align}\label{eq:imp_action}
S_\mathrm{imp} =-&\int_0^\beta d\tau\int_0^\beta d\tau'\bar{\mathbf{f}}(\tau)\mathbfcal{G}_0^{-1}(\tau-\tau')\mathbf{f}(\tau') \nonumber\\
& + U\int_0^\beta d\tau \bar{f}_{\uparrow}(\tau)\bar{f}_{\downarrow}(\tau)f_{\downarrow}(\tau)f_{\uparrow}(\tau),
\end{align}
where $\mathbfcal{G}_0$ is the so-called Weiss field, and is to be determined self-consistently to satisfy the condition
\begin{equation}\label{eq:selfcon}
\mathbf{G}_f=\mathbf{G}_\mathrm{imp} .
\end{equation}
Here, $\mathbf{G}_\mathrm{imp}$ is the Green's function of the single impurity problem \eqref{eq:imp_action}
\begin{equation}\label{eq:Dyson}
\mathbf{G}_{\mathrm{imp}}^{-1}(i\wn)=\mathbfcal{G}_0^{-1}(i\wn)-\mathbf{\Sigma}_{\mathrm{imp}}(i\wn),
\end{equation}
whereas $\mathbf{G}_f$ is the local Green's function of the lattice in the $f$-sector, cf. Eq.~\eqref{eq:def_Gk}
\begin{equation}
\mathbf{G}_f(i\wn)=\sum_\veck \mathbf{G}_{f,\veck}(i\wn)
\end{equation}
The lattice self-energy \eqref{eq:Sigma} needed to calculate $\mathbf{G}_{f,\veck}(i\wn)$ through Eq.~\eqref{eq:Gk} is approximated as
\begin{equation}
\mathbf{\Sigma}_\veck\rightarrow\begin{bmatrix}
	\mathbf{0} & \mathbf{0} \\
	\mathbf{0} & \mathbf{\Sigma}_\mathrm{imp}
\end{bmatrix}.
\end{equation}  

We satisfy the DMFT self-consistency condition by the standard iterative forward-substitution algorithm. A single DMFT iteration proceeds as follows: (i) given the $\mathbf{\Sigma}_\mathrm{imp}$ and $\Dbcs$ from the previous iteration, get new $\mathbf{G}$ using \eqref{eq:Gk}; (ii) from $\mathcal{F}_c$ determine $\Dbcs$ using \eqref{eq:order_param}. (iii) With the updated $\Dbcs$ determine a new $\mathbf{G}$ using \eqref{eq:H0},\eqref{eq:G0},\eqref{eq:Gk}. (iv) update $\mathbfcal{G}_0$ cf. $\mathbfcal{G}_0^{-1}(i\wn)=\mathbf{G}_f^{-1}(i\wn)+\mathbf{\Sigma}_\mathrm{imp}(i\wn)$. (v) given the $\mathbfcal{G}_0$ solve \eqref{eq:imp_action} to calculate $\mathbf{\Sigma}_\mathrm{imp}$. The last step is performed using Cthyb QMC impurity solver.

We note that steps (ii)-(iii) may be performed in two ways: either determine $\Dbcs$ self consistently for given $\mathbf{\Sigma}_\mathrm{imp}$, or make a single $\Dbcs$ update and solve the BCS problem in parallel with the DMFT problem. We have opted for the latter approach due to better convergence of the problem in the vicinity of phase boundaries. 

We will solve the equations on the square and the Bethe lattice. The self-consistency equations slightly simplify on the Bethe lattice and they are shown in Appendix \ref{app:bethe}. 
We initially considered the Bethe lattice, but then switched to the square lattice in order to make a better connection with the dual model solution.
Since we ignore the inter-site correlations the results on the square and Bethe lattice are similar.

We note that, on the level of a single impurity in a given SC bath, we benchmarked the Cthyb QMC calculation with the numerical renormalization group (NRG) \cite{Zitko_PRB2009} and exact diagonalization (ED) \cite{Civelli_PRL2008,Civelli_PRB2009}. The agreement was excellent. However, the application of the NRG impurity solver, would require additional programming outside the scope of this work, while ED is limited by finite number of poles on the real frequency axis. Therefore, we opted for the numerically exact QMC impurity solver.

\subsection{Dual model}

We also devise and solve  a non-interacting model that is approximately dual to model Eq.~\eqref{eq:action}: it exactly reproduces certain limits and interpolates between them. The ability of the dual model to reproduce the reentrant behavior as observed in the DMFT solution, will be a strong indication that the higher-order correlations captured by DMFT do not play an important role. In addition, the dual model solution will give us insight into the spectral functions.

To motivate the specific form of our non-interacting dual model, we start by noting that there are several limits in which the self-consistent model $H[\Dbcs]$ reduces to a clean BCS superconductor with decoupled atomic impurities. Such is the case for $V\rightarrow0$, $g\rightarrow\infty$ and/or $\epsilon_f\rightarrow \pm \infty$. In the particle-hole symmetric case $U\rightarrow\infty$ reproduces this case as well. 

Next, we observe that for an isolated Hubbard atom one can write down an exactly dual non-interacting model which reproduces the full Green's function of the original model, but not the higher-order correlation functions. This non-interacting dual model features two non-interacting orbitals connected by an appropriate hopping. One of the orbitals is dual to the original Hubbard atom, while the other can be considered a ``hidden fermion'' state\cite{Sakai_PRB2016,Sakai_PRL2016}. The coupling to the hidden fermion state plays the role of the self-energy for the dual orbital. For the derivation of the non-interacting dual model in the atomic limit, see Appendix \ref{app:dual}.

We now perform a straightforward generalization of the Hubbard atom dual model. In simple terms, we take the non-interacting part of $H[\Dbcs]$ and couple a hidden fermion state $F$ to each $f$-orbital, so that each pair $f-F$ on their own is the exact dual model to the atomic limit. Then, we introduce a copy $C$ of the $c$-band and attach it to the $F$ states in such a way that at particle-hole symmetry the hidden states $C$ and $F$ become equivalent to the dual states $c$ and $f$. This model reproduces exactly the Green's function of the model Eq.~\eqref{eq:action} in the non-interacting limit ($U=0$) as well as in all the aforementioned limits where the $f$-orbitals remain effectively decoupled from the $c$-band. This model reads
\begin{align}\label{eq:hamiltonian_dual}
H&_\mathrm{dual}[\Dbcs,n_{f\sigma}] = H^{\mathrm{HF}}_{0}[\Dbcs]\nonumber\\
&-\sum_{\veck\sigma}\xi_\veck(C^\dagger_{\veck\sigma}C_{\veck\sigma} + \mathrm{H. c.}) - \Dbcs\sum_\veck \big(C^\dagger_{\veck\uparrow}C^\dagger_{-\veck\downarrow} + \mathrm{H.c.}\big) \nonumber\\
& +V\sum_{\veck\sigma}(C^\dagger_{\veck\sigma}F_{\veck\sigma} + \mathrm{H. c.}) + (\mu + U(n_{f\sigma}-1))\sum_{\veck\sigma}F^\dagger_{\veck\sigma}F_{\veck\sigma} \nonumber\\
& +\sqrt{U^2n_{f\sigma}(1-n_{f\sigma})}\sum_\veck\big(f^\dagger_{\veck\uparrow}F^\dagger_{-\veck\downarrow} + F^\dagger_{\veck\uparrow}f^\dagger_{-\veck\downarrow} + \mathrm{H.c.}\big)
\end{align}
where $H^\mathrm{HF}_0[\Dbcs]$ is the reduced Hamiltonian introduced previously, without the repulsive interaction term, and with a Hartree-shifted $f$-level energy $\epsilon_f\rightarrow\epsilon_f+Un_{f\sigma}$.
The model is self-consistently solved for the $f$-level occupation number per spin $n_{f\sigma}\in[0,1]$.
The problem reduces to a BCS theory in an $8$-dimensional orbital/Nambu space.
See Appendix \ref{app:dual} for details.

Additionally, the dual model will allow us to gain insight in the band structure at finite $U$. Our DMFT calculation is performed in Matsubara formalism, thus one needs the ill-defined analytical continuation to obtain the spectral function. We tried the analytical continuation with the Maximum entropy method, but this resulted in the absence of any sharp features from the spectra. Thus we restrict to the dual model results when considering the electronic dispersions and local density of states.

\section{Results}

We present the results in two distinct cases: for parameters which correspond to the Kondo lattice limit of the Anderson lattice model ($n_f \approx 1$ and small double occupancy of $f$ orbitals) and away from the Kondo limit where the occupation of $f$-electrons deviates significantly from half-filling. To understand the result better, we have also solved the non-interacting $U=0$ model and the effective dual model.

\subsection{Reentrant superconductivity in the Kondo lattice limit}

\begin{figure}[t]
\centering
\includegraphics[width = 0.45\textwidth]{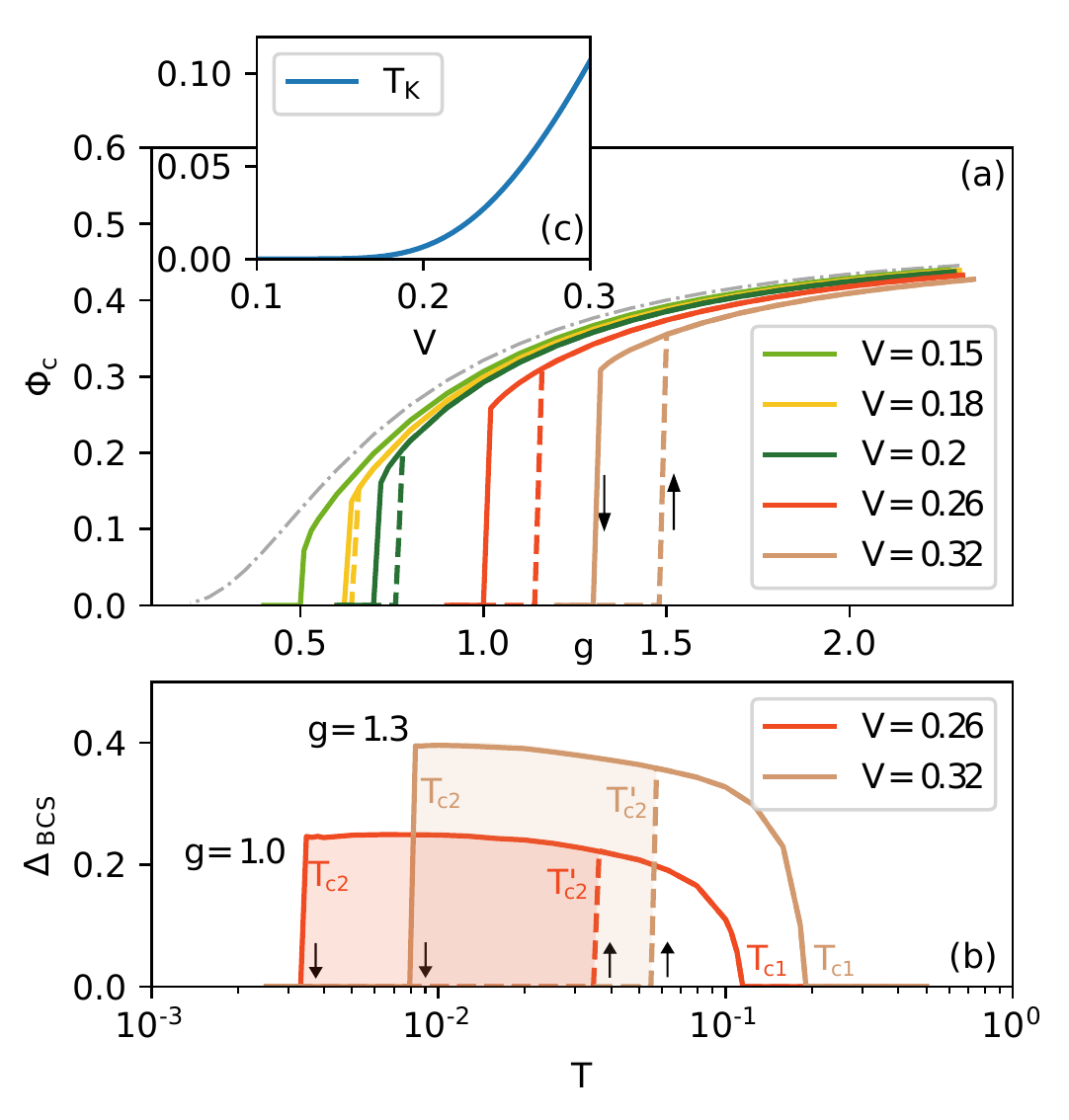}
\caption{(a) Pairing amplitude $\Phi_c$ as a function of coupling constant $g$ for several values of $V$. The gray line represents the $V=0$ BCS result. There is a discontinuous transition from the normal into the superconducting phase accompanied by hysteresis. (b) Superconducting gap as a function of temperature.
The reentrant superconducting phase appears in a broad temperature range accompanied also by a hysteresis.
(c) Inset shows the estimate of the Kondo temperature.}
\label{Kondo_limit_diagram}
\end{figure}

We study first the superconductivity for model parameters which correspond to the local moment regime, i.e~to the limit of the Kondo lattice\cite{Tanaskovic_PRB2011}. We set $\epsilon_f=-0.4$, $U=1.2$ and $\mu =0.03$ in the energy units $D=1$, and take the semicircular density of states for $c$-electrons corresponding to the Bethe lattice. These parameters give $n_{f,\sigma} \approx 0.5$ and total occupation $\sum_\sigma(n_{c,\sigma}+n_{f,\sigma}) \approx 1.9$. 
We solve the model for different values of hybridization $V$ and pairing parameter $g$. Fig.~\ref{Kondo_limit_diagram}(a) shows the pairing amplitude of $c$-electrons $\Phi_c = \langle c_{i\downarrow} c_{i\uparrow} \rangle$ as a function of the coupling $g$ at temperature $T=0.0025$. At large coupling $\Phi_c$ approaches to the single band BCS result, indicated with the {dashed-dotted} line. Transition to the superconducting phase is accompanied with hysteresis as a function of $g$: as the BCS interaction $g$ increases there is a discontinuous transition to the SC phase at $g=g_{c2}$, while as $g$ decreases the normal phase is entered at a $g_{c1} < g_{c2}$. For weaker hybridization the SC solution appears for smaller values of coupling $g$ while the hysteresis region gradually shrinks. 
We note that we did not find any indication of unconventional $s$-wave superconductivity without $g$ coupling, which was presented in Ref.~\onlinecite{BodensiekPRL2013}. This type of SC solution was not found in Refs.~\onlinecite{Wu_PRX2014,Otsuki2015,Asadzadeh13,LenzPRB2017} either.

A pronounced feature of this model is the reentrant superconductivity that we find for strong coupling $g$, Fig.~\ref{Kondo_limit_diagram}(b). At the critical temperature $T_{c1}$ there is a continuous transition to the SC phase. With decreasing temperature the SC phase persists until $T_{c2}$, where a first-order transition to the normal phase takes place. There is also a hysteresis in temperature since with the increase of $T$ the SC phase appears at $T_{c2}' > T_{c2}$.
The reentrant superconductivity resembles to what is found for diluted impurities, but a direct connection is difficult to confirm since we cannot reach very small temperatures and hence we are restricted to large $g$.

At temperatures $T \ll T_K $ the impurity spins are screened and the Fermi liquid is formed from composite heavy quasiparticles. Fig.~\ref{Kondo_limit_diagram}(c) shows the estimate of the Kondo temperature $T_K \sim e^{-1/(2\rho_0 J_K)}$. Here $\rho_0$ is the density of states of bare $c$-electrons at the Fermi level and $J_K = (\frac{1}{|\epsilon_f - \mu|} + \frac{1}{|U+\epsilon_f - \mu|})V^2$ is the Kondo coupling. One may assume that the formation of coherent quasiparticles will facilitate the superconductivity for smaller coupling $g$. However, for example for $V=0.26$ we have $T=0.0025 \ll T_K \sim 0.05$, but the critical $g$ coupling is large.  
In order to understand better why the superconductivity is so sensitive to the presence of the second band of $f$-electrons, we consider next the solution of the model in the non-interacting $U=0$ case.

\subsection{Superconductivity in the $U=0$ limit}\label{sec:U0}

\begin{figure}[t]
\centering
\includegraphics[width = 0.45\textwidth]{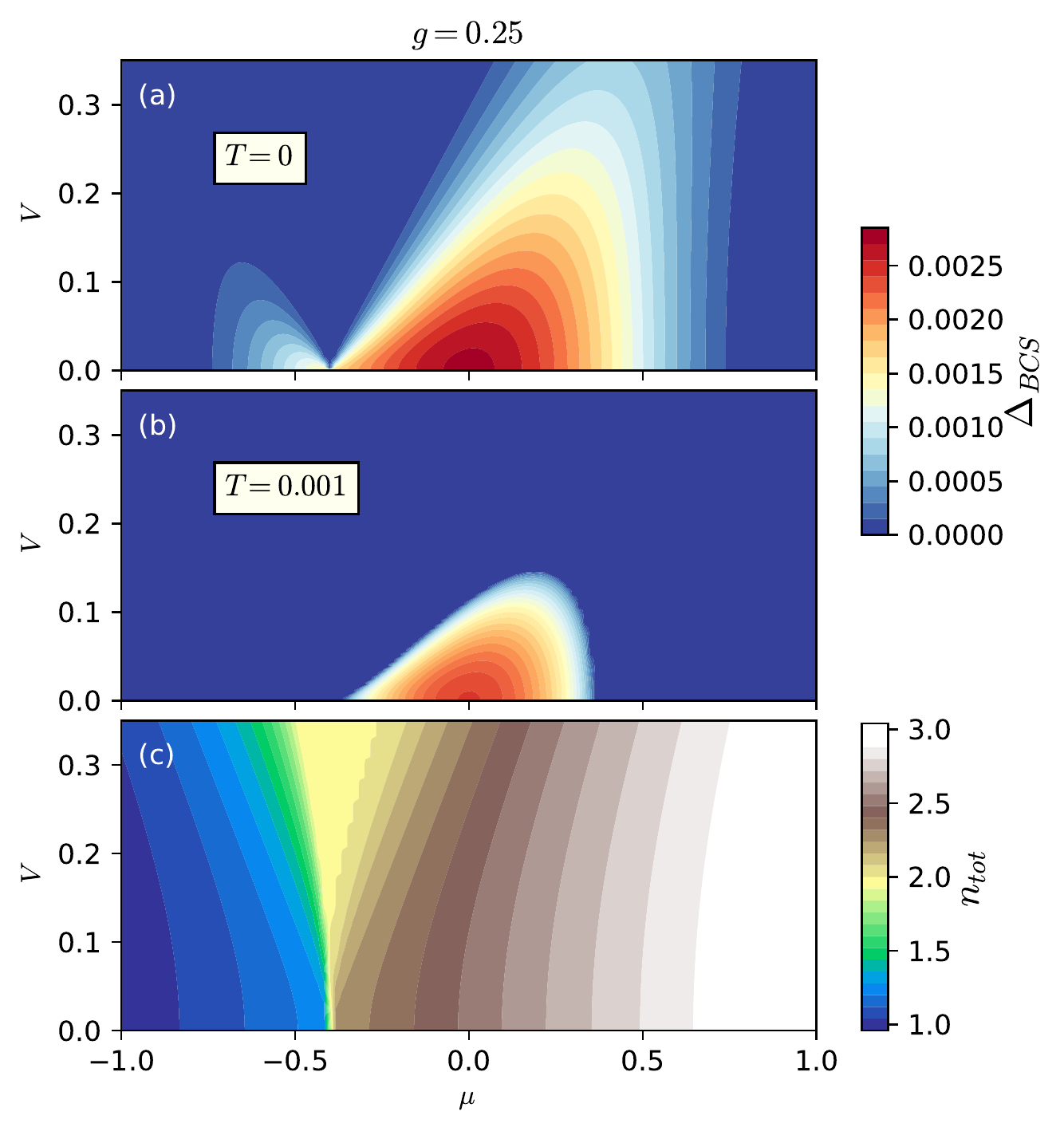}
\caption{Superconducting gap in the $V-\mu$ plane for the non-interacting ($U=0$) model at $T=0$ (a) and $T=0.001$ (b). Here $g=0.25$ and  $\epsilon_f=-0.4-\mu$. The total occupation number is shown in panel (c).
}
\label{V0_phase_diagram}
\end{figure}

In the non-interacting $U=0$ case we have derived an analytical expression for the free energy and the gap equation, see Appendix \ref{app:gap_equation}. A numerical solution of the gap equation \eqref{eq:gap_equation} is shown as a color plot on the $V - \mu$ phase diagram at $T=0$ in Fig.~\ref{V0_phase_diagram}(a) and at $T=0.001$ in Fig.~\ref{V0_phase_diagram}(b). The attractive Hubbard coupling $g$ is set to 0.25 which gives $T_{c0}\sim0.002 \ll D$ for $V=0$, while $\epsilon_f$ was kept to $-0.4-\mu$. The occupation number is varied by the chemical potential, Fig.~\ref{V0_phase_diagram}(c). We observe that the pairing amplitude is quickly suppressed by increasing the hybridization. The critical temperature $T_c$ also strongly depends on the occupation number and it goes to zero at half-filling when hybridization opens the band gap. 

This phase diagram can be understood from a simple approximate formula for $T_c(V)$ in the weak coupling limit whose derivation we now sketch. We first note that the hybridized band crosses the Fermi level at $\epsilon = \frac{V^2}{\epsilon_f-\mu} +\mu$. Then we look for the contribution of the $c$ and $f$ electrons to the hybridized eigenstate at the Fermi level. It is easy to check that the contribution of the $c$-electron is equal to $\frac{(\epsilon_f -\mu)^2}{V^2+(\epsilon_f - \mu)^2} $. Hence, the hybridized eigenstate is predominantly made of $c$-electrons for $V \ll |\epsilon_f -\mu|$, and it has mixed character for $V \sim |\epsilon_f|$. 
Then, from the usual BCS gap equation in the weak coupling limit (with the interaction cutoff set to $D=1$), $\Delta_{BCS} = 2 e^{-\frac{1}{g \rho_0}}$,
we conclude that
\begin{equation}\label{eq:approximate_bcs}
 \Delta_{BCS} = 2 e^{-\frac{V^2 + (\epsilon_f - \mu)^2}
 {g(\epsilon_f - \mu)^2  \rho [ \frac{V^2}{\epsilon_f - \mu} + \mu]}  } ,
\end{equation}
where $\rho [ \frac{V^2}{\epsilon_f - \mu} + \mu]$ is the density of states of the bare $c$-electrons at the shifted Fermi level. This expression for the superconducting gap is in excellent agreement with Fig.~\ref{V0_phase_diagram}.

\subsection{Phase diagram away from the Kondo limit}

\begin{figure}[t]
\centering
\includegraphics[width = \columnwidth]{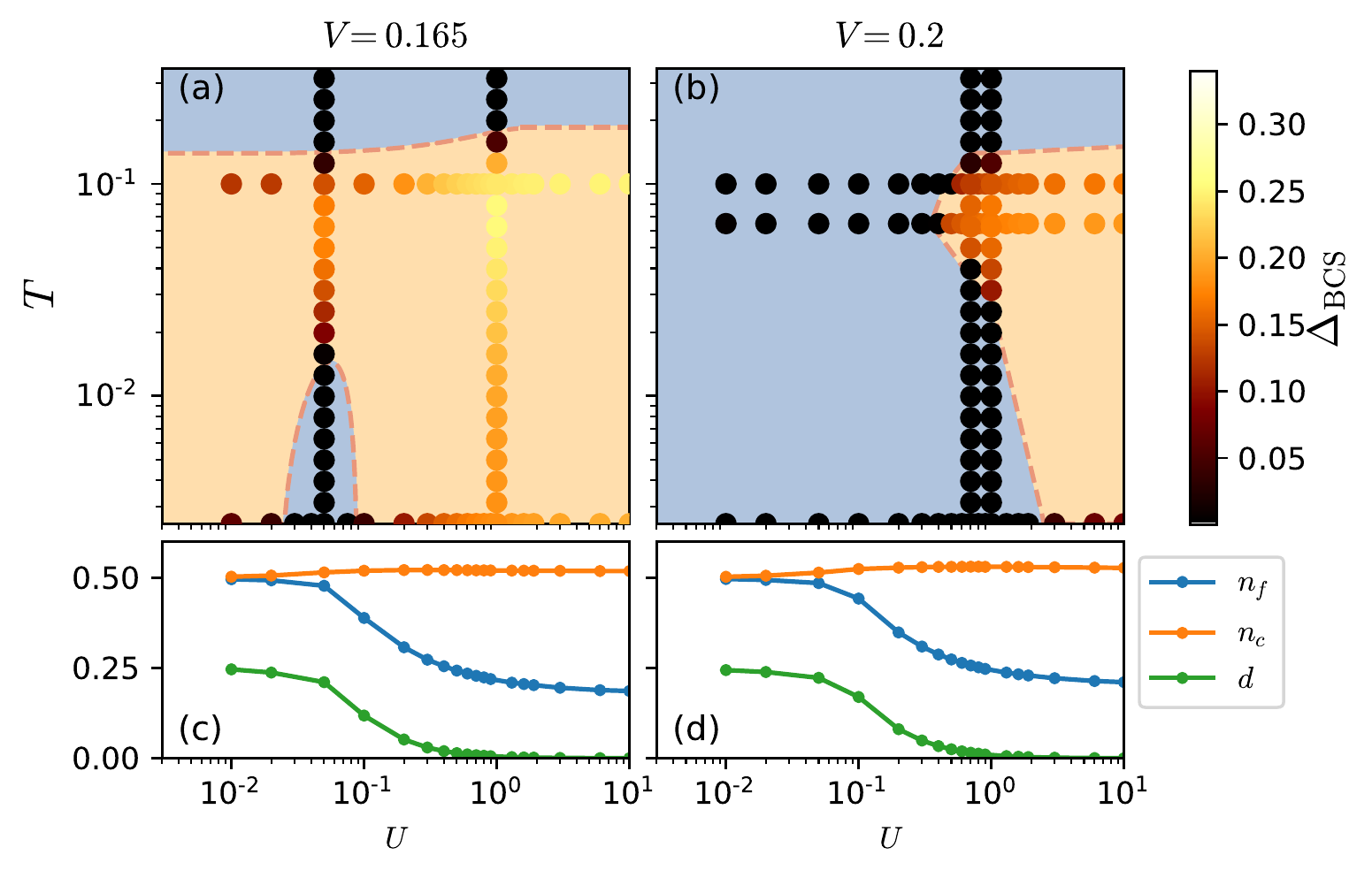}
\caption{$T-U$ phase diagram for $V=0.165$ (a) and $V=0.2$ (b). The superconducting region is shaded in beige color and the normal phase in blue. 
Colored dots are the calculated values of the superconducting order parameter at given $T$ and $U$. Here $g=1$ and $\epsilon_f=\mu=0$. Panels (c) and (d) show the corresponding occupation numbers per spin for $c$ and $f$ electrons, as well as the double occupation of the $f$ orbital.}
\label{fixedE_f_phase_diagram}
\end{figure}

In the following we probe the phase diagram of the model away from the Kondo limit at parameters accessible with our numerical methods. We show the results for the square lattice. We set $g=1$ while fixing $\epsilon_f$ and $\mu$ to 0. In Fig.~\ref{fixedE_f_phase_diagram} we present the results for $V=0.165$ (left column) and $V=0.2$ (right column). Panels (a) and (b) show the $T-U$ phase diagram. Here colored dots indicate the value of $\Delta_\mathrm{BCS}$. The beige color region indicates the SC phase, whereas the blue region corresponds to the normal phase. Panels (c) and (d) show the occupation numbers $n_c$ (orange), $n_f$ (blue) and double occupation $d$ (green) of the $f$-orbital. 

For $V=0.165$ we find that at low temperature and small $U\sim0.05$ the phase diagram exhibits an enclosed normal-phase region. Starting from this region, going up in temperature, we encounter a reentrant superconducting phase. By increasing the hybridization strength to $V=0.2$, we find that the normal phase now dominates the low-to-moderate $U$ part of the phase diagram, whereas we find reentrant superconductivity at $U\sim 1$. 

At small $U$ we are close to the non-interacting solution and we find that the superconducting phase is strongly affected by the hybridization strength, similar as in Section \ref{sec:U0}. As $U$ increases, the $f$-orbital occupation number drops and the contribution of $f$ states to the hybridized state diminishes, allowing for pairing to persist. We argue that the reentrant behavior found for $U \sim 1$ is caused by thermal excitations which reduce the hybridization at intermediate temperatures, allowing for superconductivity, before destroying the Cooper pairing at higher temperatures. In the weak coupling limit (for small $g$) we expect the phase diagram to retain these features, however with appropriately scaled $T$ and $V$.

\subsection{Dual model solution and in-gap states}

\begin{figure}[t]
\centering
\includegraphics[width = \columnwidth]{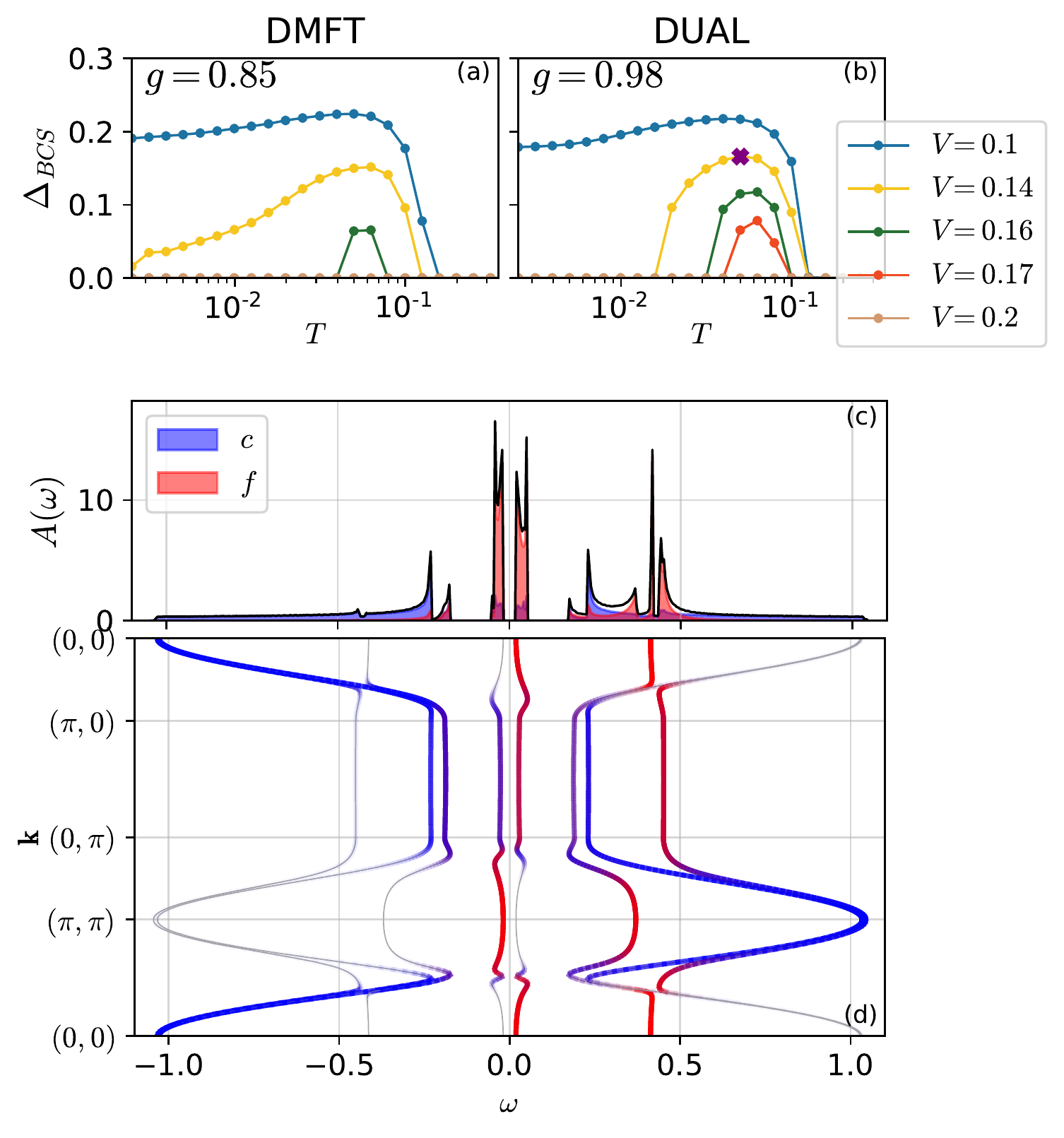}
\caption{DMFT (a) vs.~the dual model superconducting gap (b) as a function of temperature. Here $\epsilon_f=0$, $\mu=0$ and $U=0.4$. Spectral function and the dispersion relations of the $c$- and $f$-electrons in the dual model are shown in panels (c) and (d), for parameters $\Delta_\mathrm{BCS}=0.166$, $V=0.14$ which correspond to the solution indicated with purple cross in panel (b).}
\label{fig:DMFTvsDUAL}
\end{figure}

Numerical DMFT results indicate that the reentrant superconductivity that we have observed is mainly the consequence of band physics and not the consequence of physics related directly to the Kondo effect. To strengthen this argument, we consider an approximate non-interacting model which is ``dual'' to the one given by Eq.~\eqref{eq:hamiltonian}. The dual Hamiltonian given by Eq.~\eqref{eq:hamiltonian_dual} is an approximation to model Eq.~\eqref{eq:hamiltonian} which in several limits coincides with the exact solution. See Appendix \ref{app:dual} for details. The dual model also gives access to the real-frequency data and the spectral function.

In Fig.~\ref{fig:DMFTvsDUAL} we present a comparison of the DMFT superconducting gap (a) and the dual model superconducting gap (b) at $\epsilon_f=0$, $\mu=0$ and $U=0.4$. We find that the dual model approximately reproduces the DMFT results, in particular capturing the reentrant superconducting solution.  Here we adjusted $g$ to 0.98 to make the $V=0.1$ results almost coincide.

In Fig.~\ref{fig:DMFTvsDUAL}(c) we present the $c$- and $f$-electrons spectral functions. The parameters are $\Delta_\mathrm{BCS}=0.166$, $V=0.14$ and $n_{\sigma}=0.337$, which correspond to the purple cross in panel (b). We find spectral weight within the superconducting gap, which originates predominantly from the $f$-electrons.  
The upper $f$ Hubbard band is situated around $\omega\sim U$.

\section{Conclusion}\label{sec:discussion}

In summary, we have studied how the presence of dispersionless $f$-electrons hybridized with the conduction band
influences superconductivity. We solved the periodic Anderson model with an additional attractive on-site interaction between $c$-electrons restricting to the paramagnetic phase and $s$-wave superconductivity. The superconducting pairing is treated at the static mean field level and the correlations on the $f$ orbitals are treated within DMFT using Cthyb QMC impurity solver. The DMFT equations were supplemented by the self-consistency condition for the superconducting gap.

We first solved the model in the Kondo regime ($n_f \approx 1$ and small double occupancy of $f$ orbitals). We found that a large coupling $g$ was necessary in order to stabilize the superconducting solution even for small Kondo temperature $T_K$. This indicates that the many-body correlations that lead to the Kondo effect are not crucial for understanding the superconducting solution. Suppression of the superconductivity is mainly a consequence of the single-particle physics as can be understood from the noninteracting $U=0$ limit of the model. We derived a simple formula that shows that the suppression of the superconducting gap depends on the contribution of the $f$ states to the hybridized eigenstate at the Fermi level.

We scanned the phase diagram also away from the Kondo limit where the strength of the hybridization of $f$-states and appearance of the superconducting phase can be tuned by changing the parameters $V$, $U$, and $\epsilon_f$. Better insight into the band structure is obtained from the approximate dual model whose solution interpolates between several exact limits and semiquantitatively reproduces our main results. The dual model solution features in-gap states of predominantly $f$ character.

The most prominent feature of the model is its reentrant superconductivity. Here it is observed for large coupling $g$ at temperatures accessible to Cthyb QMC impurity solver. Though the reentrant superconductivity resembles to what was found previously for diluted magnetic impurities, we were not able to relate its appearance with the ratio of $T_K$ and single band $T_{c0}$. Interestingly, we found reentrant superconductivity also away from the Kondo limit by tuning the interaction $U$, which also indicates that the reentrant superconductivity is here not the consequence of higher-order many-body correlations, but rather the consequence of thermal fluctuations which weaken the hybridization of $c$-electrons making them superconducting at intermediate temperatures.
Finally, we note that our conclusions are not directly relevant to heavy-fermion systems, mainly due to unphysically large coupling constant $g$. Future study would, therefore, need to consider finite concentration of impurities,
e.g. by using real-space DMFT and to treat the attractive interaction beyond the simplest mean-field decoupling.

\section{Acknowledgments}

We thank M.~Civelli, R.~\v Zitko, and J.~Vu\v ci\v cevi\'c for useful discussions and contributions at the early stage of this project. The authors acknowledge funding provided by the Institute of Physics Belgrade, through the grant by the Ministry of Education, Science, and Technological Development of the Republic of Serbia. Numerical simulations were performed on the PARADOX supercomputing facility at the Scientific Computing Laboratory of the Institute of Physics Belgrade.

\appendix

\section{DMFT equations for the Bethe lattice}\label{app:bethe}

On the Bethe lattice there is no translational symmetry, and we may solve the DMFT as follows. The full (interacting) Green's function in the orbital-Nambu space is given by
\begin{widetext}
\begin{align}\label{eq:G_Bethe}
\mathbf{G}(i\wn,\epsilon) &= \begin{bmatrix}
	\mathbf{G}_{c}(i\wn,\epsilon) & \mathbf{G}_{cf}(i\wn,\epsilon) \\
	\mathbf{G}_{fc}(i\wn,\epsilon) & \mathbf{G}_{f}(i\wn,\epsilon)
\end{bmatrix} \nonumber\\
 &= \begin{pmatrix}
	i\wn + \mu - \epsilon & -\Dbcs & -V & 0 \\
	-\Dbcs & i\wn - \mu + \epsilon & 0 & V \\
	-V & 0 & i\wn + \mu - \epsilon_f-\Sigma(i\wn) & -\mathcal{S}(i\wn) \\
	0 & V & -\mathcal{S}(i\wn) & i\wn - \mu + \epsilon_f +\Sigma^*(i\wn)
\end{pmatrix}^{-1}.
\end{align}
\end{widetext}
In the limit of large coordination number the noninteracting density of states of $c$-electrons is equal to $\rho_0(\epsilon)=\frac{2}{\pi}\sqrt{1-\epsilon^2}$. By integrating over the density of states we extract local quantities on the lattice, viz.
\begin{align}
\mathbf{G}(i\wn) &= \begin{bmatrix}
	\mathbf{G}_{c}(i\wn) & \mathbf{G}_{cf}(i\wn) \\
	\mathbf{G}_{fc}(i\wn) & \mathbf{G}_{f}(i\wn)
\end{bmatrix} \nonumber\\
&= \int_{-D}^{D}d\epsilon\rho_0(\epsilon)\mathbf{G}(i\wn, \epsilon) .
\end{align}
The lattice self-energy is equal to the impurity self-energy $\mathbf{\Sigma}\rightarrow\mathbf{\Sigma}_\mathrm{imp}$ and the local Green's function is equal to the impurity Green's function $\mathbf{G}\rightarrow\mathbf{G}_\mathrm{imp}$. The self-consistency condition is slightly simplified in the case of the Bethe lattice:
the hybridization bath $\mathbf{\Delta}_f$ is equal to
\begin{equation}\label{bethe_selfcon}
\mathbf{\Delta}_f(i\wn)=\mathbf{V}(i\omega_n\mathbf{1}+\boldsymbol{\mu}-\mathbf{t}\mathbf{G}_c(i\wn)\mathbf{t}-\mathbf{\Delta}_\mathrm{BCS})^{-1}\mathbf{V}.
\end{equation}
Here the boldface Hamiltonian parameters $V$, $t$, $\mu$ and $\epsilon_f$ mean that they are diagonal in Nambu space, e.g. $\mathbf{V}\equiv V\bigl( \begin{smallmatrix}1 & 0\\ 0 & -1\end{smallmatrix}\bigr)$, while $\Delta_\mathrm{BCS}$ is off-diagonal $\mathbf{\Delta}_\mathrm{BCS}\equiv \Delta_\mathrm{BCS}\bigl( \begin{smallmatrix}0 & 1\\ 1 & 0\end{smallmatrix}\bigr)$ and $\mathbf{1}=\bigl( \begin{smallmatrix}1 & 0\\ 0 & 1\end{smallmatrix}\bigr)$. The Weiss field is
\begin{equation}
\mathbfcal{G}^{-1}_0(i\wn) = i\omega_n\mathbf{1}+\boldsymbol{\mu}-\boldsymbol{\epsilon}_f-\mathbf{\Delta}_f(i\wn).
\end{equation}
The self-consistent solution is obtained by solving the impurity problem (\ref{eq:imp_action}) with the self-consistency conduction (\ref{bethe_selfcon}) using iterative procedure.

\section{Dual model}\label{app:dual}

We start by considering a Hubbard atom $d$
\begin{equation}
H=-\mu\sum_\sigma d^\dagger_\sigma d_\sigma + Ud^\dagger_\uparrow d^\dagger_\downarrow d_\downarrow d_\uparrow
\end{equation}
with chemical potential $\mu$ and interaction strength $U$, whose Green's function (for Matsubara frequencies $\omega_n$) reads
\begin{equation}
G_\sigma(i\wn) = \frac{1-n_\sigma}{i\wn+\mu} + \frac{n_\sigma}{i\wn + \mu - U}.
\end{equation}
$n_\sigma\in[0,1]$ is the occupation number of spin projection $\sigma$. Writing the Dyson equation
\begin{equation}
G^{-1}(i\wn) = G_0^{-1}(i\wn) - \Sigma(i\wn)
\end{equation}
with $G_0^{-1}(i\wn)=i\wn+\mu$ the bare propagator, we derive the Hubbard atom self-energy
\begin{equation}
\Sigma_\sigma(i\wn)=\frac{Un_\sigma(i\wn+\mu)}{i\wn + \mu +U(n_\sigma-1)}.
\end{equation}
It has the property
\begin{equation}
\Sigma_\sigma(i\wn\rightarrow\infty)=Un_\sigma\equiv\Sigma_\sigma^{\mathrm{HF}} ,
\end{equation}
which is the static Hartree-Fock shift of the chemical potential. Particle-hole symmetry is achieved for $\mu=U/2$. 

We write the self energy `beyond' Hartree-Fock as
\begin{equation}
\Sigma_\sigma^\mathrm{(HF)}(i\wn) = \Sigma_\sigma(i\wn) - \Sigma_\sigma^\mathrm{HF}(i\wn)
\end{equation}
such that after some manipulations
\begin{equation}
\Sigma_\sigma^\mathrm{(HF)}(i\wn) = \frac{-U^2n_\sigma(n_\sigma-1)}{i\wn+\mu+U(n_\sigma-1)}.
\end{equation}
Writing a hybridization function of the general form
\begin{equation}
\Delta(i\wn) = \sum_\alpha\frac{|A_\alpha|^2}{i\wn - \varepsilon_\alpha} ,
\end{equation}
where $\alpha$ is some degrees of freedom, we recognize that $\Sigma^\mathrm{(HF)}$ has the form of a hybridization function with 
\begin{equation}
A_\sigma = \pm\sqrt{U^2n_\sigma(1-n_\sigma)}
\end{equation}
and 
\begin{equation}
\varepsilon_\sigma = -\mu - U(n_\sigma-1).
\end{equation}
(At particle-hole symmetry $A_\sigma\rightarrow\pm U/2$ and $\varepsilon_\sigma\rightarrow0$.) Thus, we can write a non-interacting dual model for the Hubbard atom as follows;
\begin{align}
H_\mathrm{dual}[n_\sigma]=-&\sum_\sigma(\mu-Un_\sigma)d^\dagger_\sigma d_\sigma \\
& -\sum_\sigma(\mu+U(n_\sigma-1))D^\dagger_\sigma D_\sigma \nonumber\\
& -\sum_\sigma\Big(\sqrt{U^2n_\sigma(1-n_\sigma)}d^\dagger_\sigma D_\sigma + \mathrm{H.c.}\Big) \nonumber
\end{align}
where $D$ are the ``hidden fermion'' operators dual to $d$, and $n_\sigma$ and $\mu$ need to be determined self-consistently. 

We now establish a correspondence of the Hubbard atom operators ($d$ and $D$) for the Anderson lattice model with a spin-mixing pairing term in the $c-$band, i.e.~Hamiltonian \eqref{eq:hamiltonian}]. We may identify $d$ with $f_\uparrow$, but since we are interested in solutions for any doping, we cannot identify $D$ with $f_\downarrow$. However, for $n\rightarrow1-n$ we may identify $D$ with $f_\uparrow^\dagger$, thus $\braket{dd^\dagger}=\braket{f_\uparrow f^\dagger_\uparrow}$, but $\braket{DD^\dagger}=\braket{f^\dagger_\downarrow f_\downarrow}$ for the opposite doping. Therefore the solution is to couple the model \eqref{eq:hamiltonian} to its dual at the opposite doping, where the model has the symmetry that $\Dbcs$ is the same regardless of the `sign' of the doping (i.e. $n$ or $1-n$), 
\begin{equation}
\braket{f^\dagger_\sigma(\tau)f_\sigma(0)}[n] = \braket{f_\sigma(\tau)f^\dagger_\sigma(0)}[1-n].
\end{equation}
Thus,
\begin{equation}
\braket{f_\sigma(\tau) f_\sigma^\dagger(0)}=\braket{F_\sigma^\dagger(\tau) F_\sigma(0)} ,
\end{equation}
whereas
\begin{equation}
\braket{f_\uparrow(\tau) f_\uparrow^\dagger(0)} = \braket{f_\downarrow(\tau) f^\dagger_\downarrow(0)}
\end{equation}
(similarly for $c$ and $C$). Hence, using the spinors $\Psi_\veck=\big(c_{\veck\uparrow}\;c^\dagger_{-\veck\downarrow}\;f_{\veck\uparrow}\;f^\dagger_{-\veck\downarrow}\;C_{\veck\uparrow}\;C^\dagger_{-\veck\downarrow}\;F_{\veck\uparrow}\;F^\dagger_{-\veck\downarrow}\big)^T$, the Hamiltonian matrix in orbital-Nambu space acquires the form 
\begin{widetext}
\begin{align}\label{eq:Hdual-oN}
&H_\mathrm{dual}[\Dbcs, n] = \sum_\veck\Psi_\veck^\dagger\begin{pmatrix}
	\xi_\veck & -\Dbcs & -V & 0  & 0 & 0 & 0 & 0 \\
	-\Dbcs & -\xi_\veck & 0 & V  & 0 & 0 & 0 & 0 \\
	-V & 0 & \epsilon_{f,1} & 0      & 0 & 0 & 0 & A \\
	0 & V & 0 & -\epsilon_{f,1}      & 0 & 0 & A & 0 \\
	0 & 0 & 0 & 0 & -\xi_\veck & -\Dbcs & V & 0 \\
	0 & 0 & 0 & 0 & -\Dbcs & \xi_\veck & 0 & -V \\
	0 & 0 & 0 & A & V & 0 & \epsilon_{f,2} & 0 \\
	0 & 0 & A & 0 & 0 & -V & 0 & -\epsilon_{f,2} 
\end{pmatrix}\Psi_\veck
\end{align}
\end{widetext}
with $\epsilon_{f,1} = -\mu+\Sigma^\mathrm{HF}$ and $\epsilon_{f,2}=-\epsilon$ (spin indices are dropped). If setting $V\rightarrow0$ there are two decoupled copies of the single-band BCS problem with decoupled $f$ electrons at opposite doping. Setting $U\rightarrow0$ results in two separate copies of the non-interacting model.

\begin{widetext}

\section{$U=0$ gap equation}\label{app:gap_equation}

The $c$-electron's anomalous Green's function at $U=0$ reads [from Eq.~\eqref{eq:Gk}], after Fourier transform to Matsubara space,
\begin{equation}\label{eq:BCS-anom-sum2}
\mathcal{F}_{\veck,c}(i\wn)=\frac{-\Dbcs}{(i\omega_n+\xi_\veck-\frac{V^2}{i\omega_n+\epsilon_f})(i\omega_n-\xi_\veck-\frac{V^2}{i\omega_n-\epsilon_f})-\Dbcs^2}.
\end{equation}
Following Eq.~\eqref{eq:order_param}, at self-consistency it must follow that
\begin{equation}\label{eq:conv_cond}
1=-gT\sum_{\veck n}\frac{1}{(i\wn+\xi_\veck-\frac{V^2}{i\wn+\epsilon_f})(i\wn-\xi_\veck-\frac{V^2}{i\wn-\epsilon_f})-\Dbcs^2}.
\end{equation}
We perform the infinite Matsubara sum with the standard method of contour integration. The fraction in \eqref{eq:conv_cond} can be factored as
\begin{equation}
L_\veck(\omega)=\frac{\omega^2-\epsilon_f^2}{(\omega-\lambda_{1,\veck})\cdots(\omega-\lambda_{4,\veck})}
\end{equation}
with the $\lambda$'s the eigenenergies of the $U=0$ Hamiltonian [Eq.~\eqref{eq:hamiltonian}], viz.
\begin{equation}\label{eq:eigenenergies}
\lambda_{1\cdots4,\veck}=\pm\sqrt{\frac{a_\veck\pm b_\veck}{2}},
\end{equation}
where
\begin{align}
&a_\veck=2V^2+\Dbcs^2+\epsilon_f^2+\xi_\veck^2\\
&b_\veck=\sqrt{\big(\Dbcs^2-\epsilon_f^2+\xi_\veck^2\big)^2+ 4V^2\big(\Dbcs^2+(\epsilon_f+\xi_\veck)^2\big)}.\nonumber
\end{align}
Since there are only simple poles to consider, the integration is straightforward and follows by the sum of residues of $L_\veck(\omega)f(\omega)$ at the four eigenenergies, where $f(\omega)=\beta/(e^{\beta\omega}+1)$ the counting function. After some manipulations the gap equation for the non-interacting model follows as
\begin{equation}\label{eq:gap_equation}
1=g\sum_\veck\frac{\lambda_{-,\veck}(\lambda_{+,\veck}^2-
2\epsilon_f^2)\tanh\frac{\beta\lambda_{+,\veck}}{\sqrt{8}} - \lambda_{+,\veck}(\lambda_{-,\veck}^2-
2\epsilon_f^2)\tanh\frac{\beta\lambda_{-,\veck}}{\sqrt{8}}} {\sqrt{8}\lambda_{-,\veck}\lambda_{+,\veck}b_\veck}
\end{equation}
with $\lambda_{\pm,\veck}=\sqrt{a_\veck\pm b_\veck}$. In the limit $T\rightarrow0$, $\tanh\beta\lambda_{\pm,\veck}\rightarrow1$. 

\end{widetext}

\end{document}